\documentclass[preprint,aps,preprintnumbers,amsmath,amssymb,groupedaddress]{revtex4}

\usepackage[dvips]{graphicx}
\usepackage{dcolumn}
\usepackage{bm}
\usepackage[latin1]{inputenc}

\begin{document}
\title{Determination of the neutron star mass-radii relation using narrow-band gravitational wave detector}
\author{C.H. Lenzi$^{1*}$, M. Malheiro$^1$, R. M. Marinho$^1$, C. Provid\^encia$^2$ and G. F. Marranghello$^3$}

\email{chlenzi@ita.br}

\affiliation{$^1$Departamento de F\'isica, Instituto Tecnol\'ogico de Aerona\'utica, S\~ao Jos\'e dos Campos/SP, Brazil \\
$^2$Centro de F\'isica Te\'orica, Departamento de F\'isica, Universidade de Coimbra, Coimbra, Portugal \\
$^3$Universidade Federal do Pampa, Bag\'e/RS, Brazil}

\begin{abstract}
The direct detection of gravitational waves will provide valuable astrophysical
information about many celestial objects. The most promising
sources of gravitational waves are neutron stars and black holes.
These objects emit waves in a very wide spectrum of frequencies
determined by their quasi-normal modes oscillations. In this work
we are concerned with the information we can extract from f and
p$_I$-modes when a candidate leaves its signature in the resonant
mass detectors ALLEGRO, EXPLORER, NAUTILUS, MiniGrail and
SCHENBERG. Using the empirical equations, that relate the
gravitational wave frequency and damping time with the mass and
radii of the source, we have calculated the radii of the stars for
a given  interval of masses $M$ in the range of frequencies that
include the bandwidth of all resonant mass detectors. With these
values we obtain diagrams of mass-radii for different frequencies
that allowed to determine the better candidates to future
detection taking in account the compactness of the source.
Finally, to determine which are the models of compact stars that emit
gravitational waves in the frequency band of the mass resonant
detectors, we compare the mass-radii diagrams obtained  by
different neutron stars sequences from several relativistic
hadronic equations of state (GM1, GM3, TM1, NL3) and quark matter
equations of state (NJL, MTI bag model). We verify that quark
stars obtained from MIT bag model with bag constant equal to 170
MeV and quark of matter in color-superconductivity phase are the best
candidates for mass resonant detectors.
\end{abstract}

\maketitle

\section{Introduction}

The detection of gravitational waves (GWs) will have many implications
in Physics and Astrophysics. Besides the confirmation of the
general relativity theory, it will allow the investigation of
several astrophysical phenomena, such as the existence of black
holes and the mass and abundance of neutron stars, thus opening
new scientific frontiers. The most promising sources for detection
of GWs are neutron stars  and black
holes. These objects emit waves in a very wide spectrum of frequencies determined
by their quasi-normal modes oscillations \cite{Marranghello}.

With the goal to analyze the possibility of a future detection of
the quasi-normal modes of compact stars by resonant mass detectors
(RMDs), we focalize our attention in the region of the spectrum in
the range at 0.8-3.4 kHz, which is the operation region of the
antennas: ALLEGRO, EXPLORER, NAUTILUS, AURIGA \cite{astone2007},
SCHENBERG, and MiniGrail. In particular we will work with the
frequency band of the spherical detectors SCHENBERG and MiniGrail
(2.8-3.4 kHz) \cite{minigrail2007,costa2008}.

The SCHENBERG is the second spherical detector ever built in the world and the first equipped with a set
of parametric transducers, which is installed at the Physics
Institute of the University of Sao Paulo (at Sao Paulo city,
Brazil). It has undergone its first test run in September 8, 2006,
with three transducers operational. Recent information on the
present status of this detector can be found in reference \cite{cqg2008}.
It is worth stressing also that among all known GW detectors, the spherical ones are the only one
capable to determine the direction of the incoming wave \cite{lenzi20081,lenzi20082}.

In this paper we present an extension of the work made by
Marranghello \cite{Marranghello}, where the authors show results
for a restricted band of frequency that include the SCHENBERG and
MiniGrail bandwidth. In the present work we analyze the case of a
possible future detection made by all resonant antennas and
compare the mass and radius range obtained from the frequency
bands of the GWs modes to the mass and radius calculated with
several relativistic equations of state (EoS) models.

In section 2 we introduce the f and p$_I$-mode, in section 3 we
show the mass-radii diagrams for the f and p$_I$ modes and compare them
with the ones obtained with different relativistic EoS models, in
section 4 we introduce the damping time of the f-mode and its
mass-radii diagram and finally, in section 5, we made the last
considerations.

\section{The quasi-normal modes: f and p$_I$-modes}

The neutron stars have a rich spectrum of frequencies because the
fluid perturbation oscillates in many different modes. From the GW
point of view the most important quasi-normal modes are the
fundamental mode of the fluid oscillation (f-mode), the first
pressure mode (p$_{I}$-mode), the first GW mode (w$_{I}$-mode)
\cite{kokkotas92} and the r-modes that, under certain
circumstances, can be an important source of GWs
\cite{kokkotas99}.

In this work we concentrate in the f and p$_{I}$-modes. The
fundamental mode can be described by the density distribution
inside the star, while the p-mode is the pressure restoration
force. In reference \cite{Benhar2004}, the authors have obtained
an empirical formulae for the frequencies of these two modes as a
function of the mass and radius using a wide sample of equations
of state:
\begin{equation}\label{eq1}
  \nu_f = (0.79\pm 0.09) +(33\pm 2)\sqrt{\frac{M}{R^3}},
\end{equation}
\begin{equation}\label{eq2}
 \nu_p = \frac{1}{M}\left[(-1.5\pm 0.8) + (79\pm 4)\frac{M}{R}\right],
\end{equation}
where the mass and the radii are given in km (remember that
$M_\odot \thickapprox 1.477$ km), while $\nu_f$ and $\nu_p$ are
given in kHz. Using the empirical relations (\ref{eq1}) and
(\ref{eq2}), we have calculated the radii $R$ of the stars for a
given interval of masses $M$ in the range of frequencies that
include the bandwidth (0.8-3.4 kHz) of all RMDs in operation. In
Table (1) we can see the resonant frequencies of these detectors.

\begin{table}[ht]
\begin{center}
\caption{Frequency band of the RMDs in operantion in the world.}
\begin{tabular}{llcc}  \hline \hline
Antena    & Location    &   Freq.(Hz)  &   Type  \\ \hline

ALLEGRO \ \ \ \ \ \ \ & Baton Rouge &         890-920          & Bar       \\
 EXPLORER             &  CERN       &         895-920          & Bar       \\
NAUTILUS              & Frascai     &         905-925          & Bar       \\
 AURIGA               & Legnaro     &         850-930          & Bar       \\
SCHENBERG             & S\~ao Paulo &  \ \ \ 3100-3300 \ \ \   & Spherical \\
MiniGrail             &    Leiden   &        2800-3000         & Spherical \\
\hline
\end{tabular}\label{table1}
\end{center}
\end{table}

Through these relations we have obtained diagrams for p$_I$ and
f-modes that relate GW frequency with masses and radii of the
sources. These diagrams allow us to determine the better
candidates for a future detection by resonant antennas from the
compactness of the star. We can see in figure (1) and (2) the f
and p$_I$-mode's mass-radii diagrams where the different gray
scale identify the different frequencies.

\section{Comparison of the mass-radius diagrams with the ones
obtained by relativistic EoS models }

To determine what relativistic models of compact stars emit GW in
the frequency bands of the RMDs we compare the diagrams of the
relations (\ref{eq1}) and (\ref{eq2}) with some neutron stars
masses and radii sequences obtained by different relativistic
models that generate several equations of state for hadronic
matter such as models NP, NPH, NPHQ with and without isovector-scalar $\delta$ \cite{Constanca01}, namely
\begin{enumerate}
    \item[-] the models with parameters set GM1, GM3, NL3, TM1 \cite{Constanca02,Constanca03}
\end{enumerate}
 and some for quark strange matter as 
\begin{enumerate}
\item[-] Nambu-Jona-Las\'inio model (NJL) \cite{Constanca4}, color-flavor locked phase (CFL) \cite{Sanjay} and the MIT bag model to different values of the bag constant \cite{MITbag} and hybrid star EoS.
\end{enumerate}

Relativistic hadronic models have been widely used in order to
describe nuclear matter, finite nuclei, stellar matter properties,
and recently in the high temperature regime produced in heavy ion
collisions \cite{debora}. Many variations of the well known
quantum hadrodynamic model \cite{sw} have been developed and used
along the last decades. Some of them rely on density dependent
couplings between the baryons and the mesons
\cite{original1,original2,tw,br,gaitanos,twring1,twring2} while others use constant
couplings \cite{nl3,tm1,glen}. Still another possibility of
including density dependence on the lagrangian density is through
derivative couplings among mesons and baryons
\cite{delf1,delf2,chiappa1} or the coupling of the mediator mesons
among themselves \cite{nlwr1,nlwr2,nlwr3}. The relativistic model couplings are
adjusted in order to fit expected nuclei properties such as
binding energy, saturation density, compressibility and energy
symmetry at saturation density, particle energy levels, etc. These
same relativistic models are extrapolated to higher densities as
in stellar matter and the results obtained for the neutron star
masses and radii are quite good in comparison with the
astronomical observations. In the case of bare quark stars, the
strange matter inside the star is usually describe by the MIT bag
model, a Fermi gas of free quarks with a vacuum energy known as
the bag constant, or by chiral models like the Nambu-Jona-Las\'inio
(NJL) model that has a dynamical chiral symmetry breaking
mechanism that originates mass for the quarks. Recently, the
possibility that quarks can be paired at high densities and be in
a color superconductive phase has originated new quark matter
equations of state, that depending on the pairing interaction, can
be quite stiff and produce large stars masses and radii
\cite{MM1,MM2,MM3}. The main feature of a quark star, since they
are bound by the strong force and not by gravity, is that they
are more compact and have smaller mass to radius ratio than a
neutron star. As we will see, it is this fact that strange stars can
have small radii will explain the high frequency GWs modes
produced by this type of stars.

We can see in the diagrams (1) and (2) that the frequency band of
the RMDs are on the dark region, where it is expected that GWs generate
from less compact neutron stars in both diagrams. This fact shows
the impossibility of a future detection,  by cylindrical antennas,
of relativistic neutron star candidates emitting gravitational
wave on p$_I$ or f-mode. However, we can see on the diagrams that
for the spherical detectors bandwidth, MiniGrail and Schenberg,
have some candidates near their resonant frequencies. The most
probable source would correspond to a very compact object with
radius smaller than 10 km. The models that fulfill this condition
are models of strange quark stars, as preview in
\cite{Marranghello}. This fact is confirmed when we compare with
the compact star sequence generated from the MIT bag model (with
bag constant $B^{1/4} = 170$ MeV), NJL model and CFL of quark matter. On the other hand the p$_I$-mode would
only be expected to come from less compact neutron stars.

\section{The damping time}

How can we distinguish the f-mode in a putative detection? And how
to determine the mass and radii of the star? The damping time is
the response for these questions \cite{Marranghello}. In
\cite{Benhar2004} the authors obtained an empirical relation for
the f-mode damping time as function of the radius and mass,
described by:
\begin{equation}\label{eq3}
 \tau_f = \frac{R^4}{cM^3}\left[ (8.7 \pm 0.2)\cdot10^{-2} + (-0.271 \pm 0.009) \frac{M}{R} \right]^{-1}.
\end{equation}
Even though the RMDs can not determine the damping time properties
with small errors, we use the empirical relation (\ref{eq3}) to
calculate the damping time given the intervals of radii and mass
$(R,M)$ obtained with relation (\ref{eq1}). We can get a new
mass-radius diagram, but doing a distinction in the damping time.
We compare the diagram with models of quark stars, CFL and MIT bag model with bag constant equal to 170
MeV. Results obtained can be seen in figure (3). Through these
results we can estimate the masses and radii of the stars solving
the inverse problem, as show in \cite{kokkotas99}.

\section{Summary}

RMDs bandwidth are on the spectrum regions with a few (or none)
neutron stars candidates emiting GWs through their f and
p$_I$-modes. However, the spherical detectors are on a region
where the f-modes of very compact objects can be detected. All
sequences of neutron stars described by quark matter models are on
the region near to MiniGrail and Schenberg bandwidth, but the MIT
bag model with bag constant equal to 170 MeV and CFL of the quark matter with the bag constant 200 MeV and the gap 100 MeV are the best candidates for these
detectors, as we can see in figures (1) and (3). On the other hand
the detection of the f and p$_I$-modes of neutron stars by bar
detectors is unlikely, because their bandwidth is located in low
frequencies.

\begin{acknowledgements}
This work was partially supported by FEDER and FCT (Portugal)
under the project PDCT/FP/64707/2006. The
CHL, MM, RMM thank the financial support given by CAPES through
the fellowship 2071/07-0 and the international cooperation program
Capes-Grices between Brazil-Portugal.
\end{acknowledgements}


\begin{figure}[ht]
\begin{center}
\includegraphics[width=0.9\linewidth]{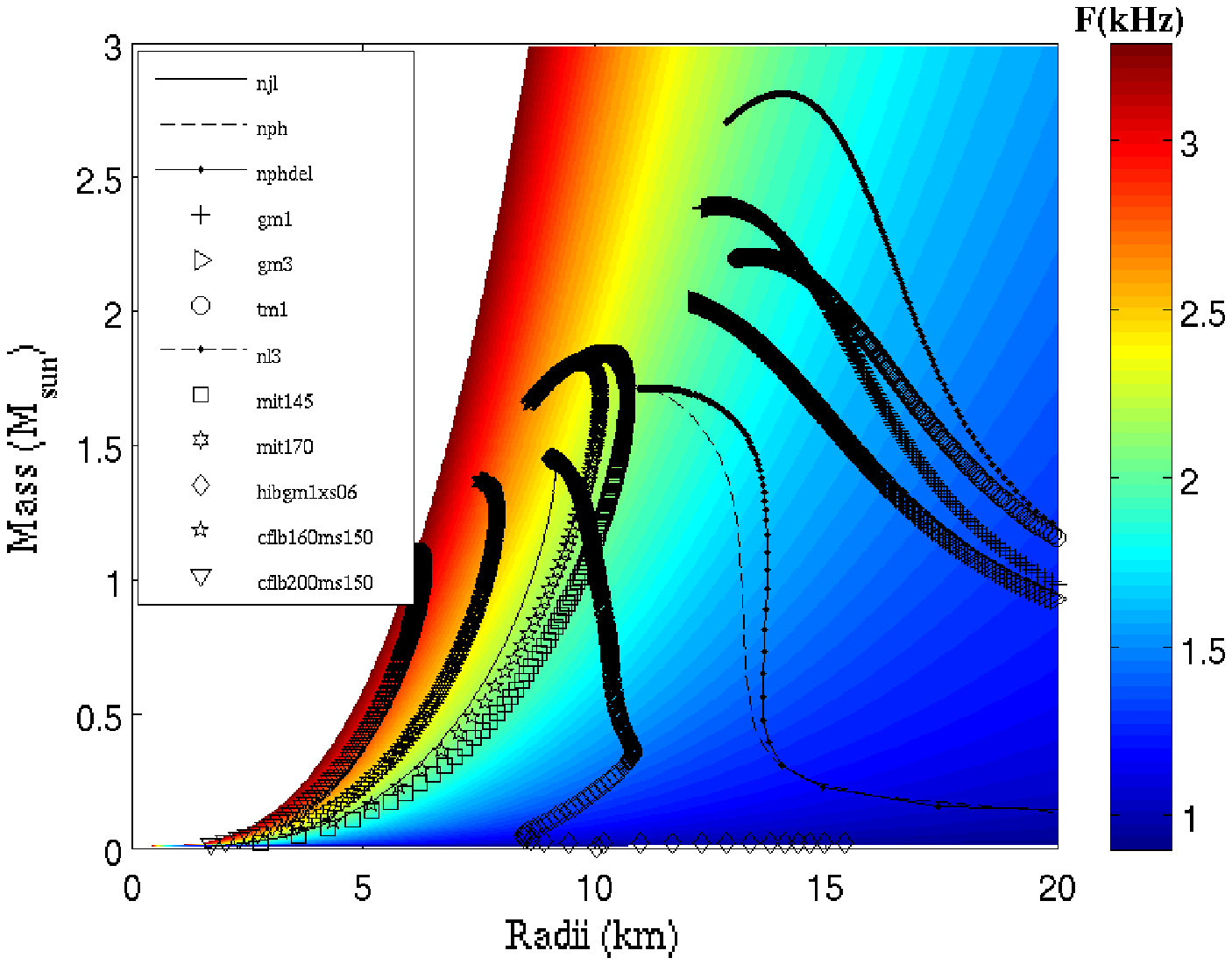}
\caption{Mass-radii diagram of the empirical relation \ref{eq1} abtained by Benhar {\it et al} for f$_I$-mode. We compare with some models of relativistic EoS. The differents gray scale identify the different frequencies.} \end{center}\label{figure1}
\end{figure}

\begin{figure}[ht]
\begin{center}
\includegraphics[width=0.9\linewidth]{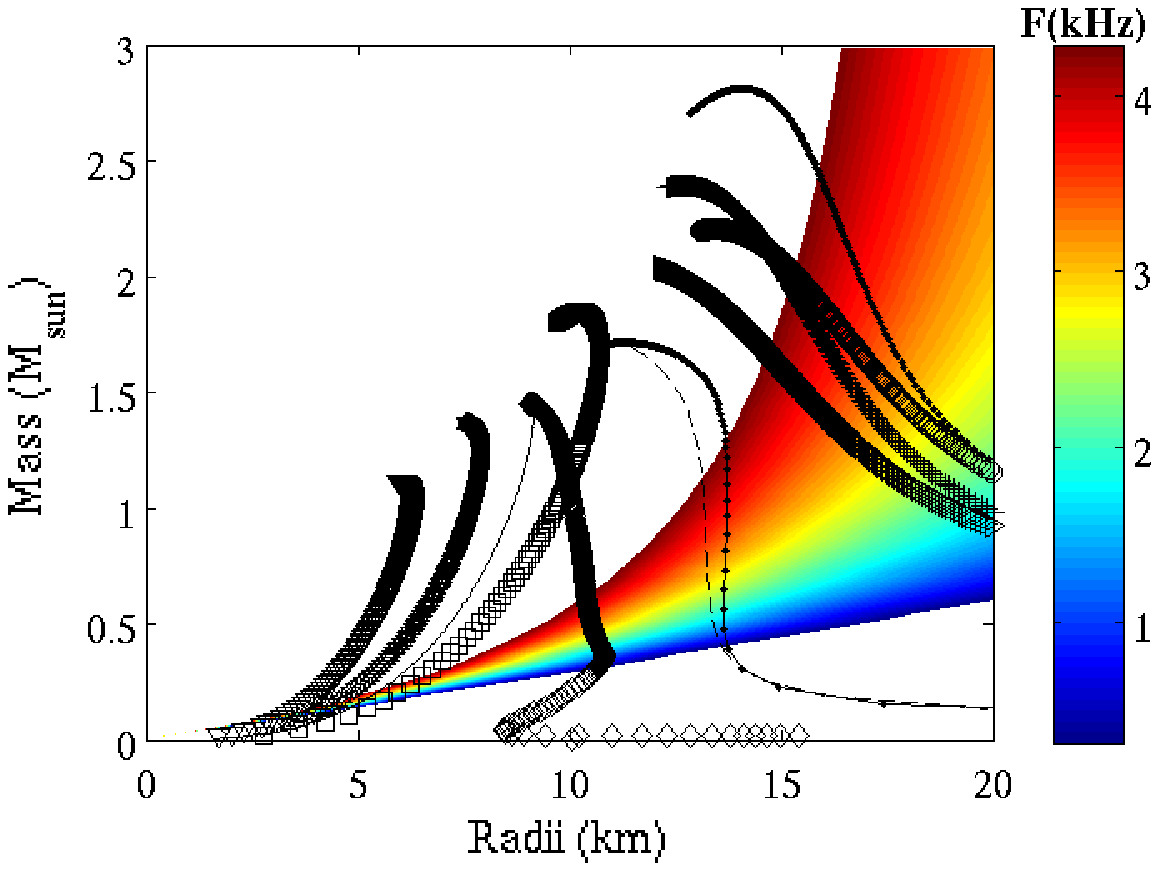}
\caption{Mass-radii diagram of the empirical relations \ref{eq2} abtained by Benhar {\it et al} for p$_I$-mode. We compare with some models of relativistic EoS. The differents gray scale identify the different frequencies.}                                                        \end{center}\label{figure2}
\end{figure}

\begin{figure}[ht]
\begin{center}
\includegraphics[width=0.9\linewidth]{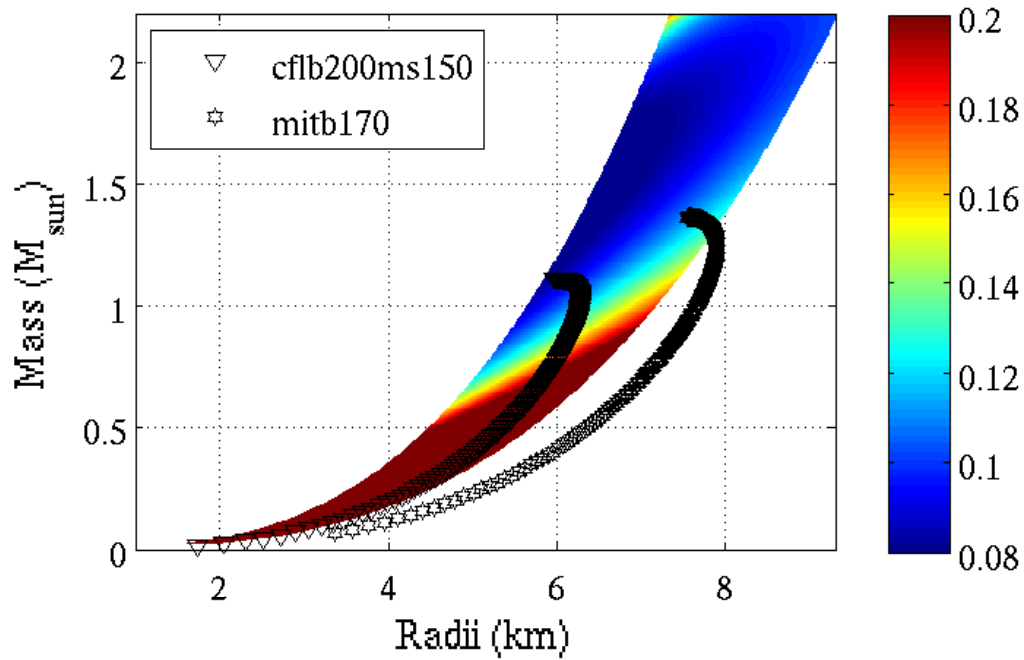}
 \caption{Mass-radii diagram of the empirical relations \ref{eq2} abtained by Benhar {\it et al} for damping time of the f$_I$-mode. We compare the diagram with models that discribe quark stars, CFL of the quark matter with bag constant 200 MeV and gap 100 MeV and MIT bag model with bag constant equal to 170 MeV.  The differents gray scale identify the different damping time.}
\end{center}\label{figure3}
\end{figure}

\end{document}